\input harvmac
\overfullrule=0pt
\Title{\vbox{
\hbox{USC-96/01}
\hbox{hep-th/9601059}
}}
{$N=2$ Super Yang-Mills and Subgroups of $SL(2,Z)$}
{\baselineskip=12pt
\centerline{Joseph A. Minahan and Dennis Nemeschansky}
\bigskip
\centerline{\sl  Department of Physics and Astronomy}
\centerline{\sl University of Southern California}
\centerline{\sl Los Angeles, CA 90089-0484}
\medskip
\bigskip
\centerline{\bf Abstract}

}

\noindent
We discuss $SL(2,Z)$ subgroups appropriate for the study of $N=2$ Super 
Yang-Mills with $N_f=2n$ flavors.  Hyperelliptic curves describing
such theories should have coefficients that are modular forms of
these subgroups.  In particular, uniqueness arguments are sufficient
to construct the $SU(3)$ curve, up to two numerical constants, which
can be fixed by making some assumptions about strong coupling behavior.
We also discuss the situation for higher groups.  We also include
a derivation of the closed form $\beta$-function for the $SU(2)$ and $SU(3)$
theories without matter, and the massless theories with $N_f=n$.   

\Date{1/96}
\vfil
\eject

\def\NP{{\it Nucl. Phys.\ }}
\def\PL{{\it Phys. Lett.\ }}

\def\PRL{{\it Phys. Rev. Lett.\ }}

\def\vth{\vartheta}

\def\th{\theta}
\def\eps{\epsilon}

\lref\SWI{N. Seiberg and E. Witten, {\bf hep-th/9407087},
        \NP{\bf 426} (1994) {19}.}
\lref\SWII{N. Seiberg and E. Witten, {\bf hep-th/9408099},
        \NP {\bf B431} (1994) {484}.}
\lref\KLTY{A. Klemm, W. Lerche, S. Theisen and S. Yankielowicz,
        {\bf hep-th/9411048}, \PL {\bf B344} (1995) {169}.}
\lref\AF{P.C. Argyres and A.E. Faraggi, {\bf hep-th/9411057},
        \PRL {\bf 73} (1995) {3931}.}
\lref\HO{A. Hanany and Y. Oz, {\it On the
	Quantum Moduli Space of Vacua of N=2 Supersymmetric $SU(N_c)$
	Gauge Theories, {\bf hep-th/9505075}, {\it Nucl. Phys.} {\bf B452} 
(1995) 283-312.}}
\lref\APS{P.~C.~Argyres, M.~R.~Plesser and A.~D.~Shapere, 
{\it The Coulomb Phase of N=2 Supersymmetric QCD}, {\bf hep-th/9505100},
{\it Phys. Rev. Lett.} {\bf 75} (1995) 1699-1702.} 
\lref\FK{H.~M.~Farkas and I.~Kra, {\it Riemann Surfaces}, Springer-Verlag 
(1980), New York.}
\lref\Clemens{C.~H.~Clemens, {\it A Scrapbook of Complex Curve Theory}, Plenum
Press (1980), New York.}
\lref\MN{J. Minahan and D. Nemeschansky, 
{\it Hyperelliptic curves for Supersymmetric Yang-Mills}, 
{\bf hep-th/9507032}, {\it to appear in Nucl. Phys. B}.}
\lref\AD{P. Argyres and M. Douglas, {\it New Phenomena in $SU(3)$ 
Supersymmetric Gauge Theory}, {\bf hep-th/9505062}, 
{\it Nucl. Phys.} {\bf B448} (1995) 93-126}
\lref\AY{O. Aharony and S. Yankielowicz, {\it Exact Electric-Magnetic Duality 
in $N=2$ Supersymmetric QCD Theories}, {\bf hep-th/9601011}.}
\lref\Kob{N. Koblitz, {\it Introduction to Elliptic Curves and Modular Forms},
Springer-Verlag, (1984), New York.}
\lref\Sch{B. Schoeneberg, {\it Elliptic Modular Functions; An Introduction},
Springer-Verlag, (1974), New York.}
\lref\FP{D. Finnell and P. Pouliot, {\it Instanton Calculations Versus Exact 
Results In 4 Dimensional Susy Gauge Theories}, {\bf hep-th/9503115}.}  
\lref\DW{R. Donagi and E. Witten, {\it Supersymmetric Yang-Mills Systems And 
Integrable Systems}, {\bf  hep-th/9510101}.}
\lref\MWI{E. Martinec and N. Warner, {\it Integrable systems and 
supersymmetric gauge theory}, {\bf hep-th/9509161}.}
\lref\Mart{E. Martinec, {\it Integrable Structures in Supersymmetric Gauge 
and String Theory}, {\bf hep-th/9510204}.}
\lref\MWII{E. Martinec and N. Warner, {\it Integrability in N=2 Gauge Theory: 
A Proof}, {\bf hep-th/9511052}.}
\lref\GKMMM{A.Gorsky, I.Krichever, A.Marshakov, A.Mironov and A.Morozov,
{\it Integrability and Seiberg-Witten Exact Solution}, {\bf hep-th/9505035},
{\it Phys. Lett.} {\bf B355} (1995) 466-47.}
\lref\NT{T. Nakatsu and K. Takasaki, {\it Whitham-Toda hierarchy and $N = 2$
 supersymmetric Yang-Mills theory}, {\bf hep-th/9509162}.}
\lref\IM{H. Itoyama and A. Morozov, {\it Integrability and Seiberg-Witten 
Theory: Curves and Period}, {\bf hep-th/9511126}.}

\newsec{Introduction}

In their classic papers, Seiberg and Witten found elliptic curves that
describe the exact effective actions for $N=2$ $SU(2)$ gauge theories,
with and without matter \refs{\SWI,\SWII}.  
In the case of $N_f=0,1,2,3$, the curve has
coefficients that are holomorphic functions of the expectation values
and bare masses of the theory.  However, since this theory has a nonzero
$\beta$-function, the coefficients of the curve must depend on a scale
$\Lambda$ as well.  

But in the case where $N_f=4$, or when there is
a hypermultiplet transforming in the adjoint of $SU(2)$, then the 
$\beta$-function
is zero, and there no longer is dependence on the scale, instead there is
dependence on a dimensionless parameter $\tau$.  
In the massless case, $\tau$ can be
interpreted as the coupling of the theory.  The coefficients of the curve
turn out to be modular forms of $\tau$ under a subgroup of $SL(2,Z)$, 
$\Gamma(2)$.  

For higher gauge groups, there is a straightforward generalization of
the $SU(2)$ case for theories with nonzero 
$\beta$-functions \refs{\AF\KLTY\HO{--}\APS}, 
up to possible constant coefficients.
However, in the case where $\beta=0$, there are difficulties 
present.  We expect the curve to be described by a parameter $\tau$,
but one must be careful with its interpretation.  For instance,  a curve
was presented in \APS\ which was derived by matching it to the $SU(2)$ curve
and taking certain masses and expectation values to infinity.  In taking
this limit, one of the $U(1)$ subgroups decouples as its effective
coupling runs to zero.  What is left is the original Seiberg-Witten
theory and hence the parameter $\tau$ should be identified with the
coupling of the remaining $SU(2)$ subgroup.  The curve is written in
terms of $\Gamma(2)$ modular {\it functions}, 
reflecting the symmetries of this leftover $SU(2)$ subgroup.

 On the other hand,
 in \MN\ a curve was found for the $SU(3)$ case by starting with a period
matrix and finding the curve.  The period matrix was assumed to be a
constant $\tau$ multiplied by the Cartan matrix.  The parameter
$\tau$ is actually the true coupling when all bare masses are zero
and the expectation value $u=\langle\tr\phi^2\rangle$ satisfies $u=0$.  
The curve
was written in terms of genus two theta functions, and the symmetry
group of this curve reflects the symmetry group of the classical $SU(3)$
coupling, not the coupling for an $SU(2)$ subgroup when a $U(1)$ decouples.
As it so happens, at weak coupling
the curve in \MN\  is equivalent to the curve in \APS, once
one takes into account that the coupling parameter $\tau$ runs in going from
the massless case to the $SU(2)$ limit.  At strong coupling,
the identification of the curves becomes more problematic and basically
requires redefining one or more of the gauge invariant expectation values
as well as the bare masses.  In any case, the curve in \MN\ more fully reflects
the symmetries for an $SU(3)$ gauge theory.  Writing the curve in this
form might also assist in finding the corresponding integrable models
for $SU(n)$ with $N_f=2n$ \refs{\GKMMM\MWI\NT\DW\Mart\MWII{--}\IM}.

We will argue that the symmetry group of
the classical coupling for $SU(n)$ is the $SL(2,Z)$ subgroup  $\Gamma_1(n)$ 
($\Gamma_1(2n)$) for $n$ odd (even).  This suggests that the appropriate 
curves have coefficients that are modular forms of this subgroup.
Using results from \MN,   
we will see that the curve for $SU(3)$ is quite simple and elegant and
its similarity to the $SU(2)$ curve is rather striking.  The $SU(3)$ curve
has coefficients that are modular forms of $\Gamma_1(3)$.  The dimension
of the space of such forms is sufficiently small in order to determine
the curve up to constant coefficients.  These coefficients depend on
nonperturbative effects and can be determined by insisting on certain
behavior at strong coupling.      

At this point,
we do not yet know how to construct the curves for higher $SU(n)$, partly 
because
of the large number of modular forms for $\Gamma_1(n)$ or $\Gamma_1(2n)$.
We will make some general observations about this case that will, hopefully,
lead to a solution.

In section 2 we discuss the subgroups of $SL(2,Z)$ appropriate for the 
classical $SU(n)$ coupling.  In section 3 we review the $SU(2)$ case with
$N_f=4$.  In this section, we also include a derivation of the $N_f=0,1,2$
$\beta$-functions in closed form, 
which can be easily derived from the massless $N_f=4$
curve, and to the best of our knowledge, has not previously appeared in
the literature.  In section 4 we discuss the $SU(3)$ theory,  complete
with a derivation of its $\beta$-function for $N_f=0,3$.  In section 5
we discuss some issues for $SU(n)$, $n\ge 4$.  

\newsec{$\Gamma_1(n)$ and $\Gamma_1(2n)$}

For gauge group $SU(n)$, the classical coupling matrix is given by $T=\tau C$,
where $C$ is the matrix
\eqn\cc{
C=\left(\matrix{2&1&1&\dots&1\cr 1&2&1&\dots&1\cr\vdots&&\ddots\cr 1&\dots&1&1&2}\right)
}
and $\tau={\th\over2\pi}+{4\pi i\over g^2}$.  We have chosen the Cartan
basis to be generated by the gauge fields $A_i-A_n$.  Clearly, the
theory should be invariant under $\tau\to \tau +1$, which corresponds
to shifting $T$ by $C$.  In fact this invariance should carry over
to the true quantum coupling matrix, $T_q$, that is the theory is
 invariant under $T_q\to T_q+C$.  $T_q$ is actually the period
matrix for the hyperelliptic curve that describes the theory.  The
period matrix will appear in the curve in terms of genus $n-1$ theta
functions, which are invariant when any component of $T_q$ is
shifted by an even integer.  Hence $T_q$ is invariant under any shift
that is equal to $C$ mod 2.  

$T_q$ is also invariant under any $Sp(2n-2,Z)$ 
transformation that is conjugate to $C$ mod 2.
The
inverse of $C$ is
\eqn\cci{
C^{-1}={1\over n}\left(\matrix{n-1&-1&-1&\dots&-1
\cr -1&n-1&-1&\dots&-1\cr\vdots&&\ddots\cr -1&\dots&-1&-1&n-1}\right)
}
and hence the theory is invariant under 
\eqn\Ttrans{\eqalign{
&T_q\to T_q+nC^{-1}\qquad\qquad n\  {\rm odd}\cr
&T_q\to T_q+2nC^{-1}\qquad\qquad n\ {\rm  even}.
}}
But the theory is also invariant  under the conjugate transformation
\eqn\Ttransc{\eqalign{
&T_q\to T_q(nC^{-1}T_q+I)^{-1}\qquad\qquad n\ {\rm odd}\cr
& T_q\to T_q(2nC^{-1}T_q+I)^{-1}\qquad\qquad n\ {\rm even},
}}
where $I$ is the identity matrix. 
Let us for the moment assume that $T_q$ is $\tau C$.
Then under the transformation in \Ttransc, $T_q$ transforms to
\eqn\Ttranscc{\eqalign{
&T_q\to {\tau\over n\tau+1}C\qquad\qquad n\ {\rm  odd}\cr
& T_q\to {\tau\over 2n\tau+1}C\qquad\qquad n\  {\rm  even.}
}}
Hence we see that for the classical form of the coupling matrix,
the theory is invariant under the transformations $\tau\to\tau+1$
and $\tau\to\tau/(n\tau+1)$  (or $\tau\to\tau/(2n\tau+1)$).  These
two transformations generate a subgroup of $SL(2,Z)$, $\Gamma_1(n)$
(or $\Gamma_1(2n)$.)  The group elements of $\Gamma_1(n)$ are
\eqn\Gammat{
\left(\matrix{1&b\cr0&1}\right)\qquad{\rm mod\ }n.
}
Unlike the subgroups $\Gamma(n)$, $\Gamma_1(n)$ is not a normal
subgroup of $SL(2,Z)$, however a lot is known about the modular
forms under these groups.  (For a nice discussion, see chapter 3 of \Kob.)

Before going further, we need to stress one point.  If $n\ge4$, then
the true coupling matrix cannot be proportional to $C$.  A way to
see this is to note that $C$ is invariant under an $Sp(2n-2,Z)$ subgroup
which is isomorphic to $S_n$, the permutation group on $n$ elements.
But this would imply that the period matrix, and hence the Riemann surface
is invariant under such a group.  But if the genus is three or greater,
then a hyperelliptic surface does not have such a symmetry.  The surfaces
with an $S_n$ symmetry are constructed from $n$ sheets, and the permutation
acts by exchanging sheets, hence the surface cannot be hyperelliptic.

For the $SU(3)$ case, $T_q$ has the classical form so long as the
expectation values have a $Z_3$ symmetry.  This occurs if 
$\langle s_k\rangle=0$, $k\ne3$ and if $t_k=0$, $n\ne3,6$. 
$s_k$ is the order $k$
symmetric homogeneous
polynomial 
of the $\phi_i$, the uncharged component fields of the adjoint scalar, and
$t_k$ are the order $k$ homogeneous symmetric polynomials of the six
bare masses.

\newsec{Review of $SU(2)$ with $N_f=4$}

The $SU(2)$ coupling is of course a scalar, so the quantum coupling
has the same form as the classical coupling.  The
symmetry group is $\Gamma_1(4)$, which is the same as $\Gamma(2)$
under the rescaling $\tau\to 2\tau$.

Let us define the quantities $f_\pm(\tau)=\th_2^4(2\tau)\pm\th_1^4(2\tau)$,
where $\th_{1,2}$ are standard genus one theta functions.  $f_+$ and
$f_-$ are weight two modular forms of $\Gamma_1(4)$.  In other words,
under the transformation $\tau\to (a\tau+b)/(c\tau+d)$, with
$\left(\matrix{a&b\cr c&d}\right)$ an element of $\Gamma_1(4)$,
$f_\pm\to(c\tau+d)^2f_\pm$.  In fact, $f_+$ and $f_-$ generate
all of the even weight forms.  Also, under the transformation 
$\tau\to -1/(4\tau)$, which is not actually in $\Gamma_1(4)$, the
weight two forms transform as 
\eqn\fpmt{
f_\pm(\tau)\to \pm(4\tau)^2i^{-2}f_\pm(\tau).} 

Let us suppose that $\tau$ is the coupling when all four bare masses
are zero.  There exists an $SL(2,C)$ transformation that maps the massless
cubic curve in \SWII\ to the quartic curve
\eqn\quartic{\eqalign{
y^2&=(f_-(\tau)x^2-u)^2 +(f_+^2(\tau)-f_-^2(\tau))x^4\cr
   &=P(x)^2+(f_+^2(\tau)-f_-^2(\tau))x^4.}
}
Notice that under this 
parameterization, $y$ and $u$ have weight two under $\Gamma_1(4)$,
while $x$ has weight zero.  The period matrix for this genus one
surface is $2\tau$.  Notice further that the curve is invariant under
$\tau\to-1/(4\tau)$, since the minus sign that $f_-$ picks up can
be absorbed into $x$.  

If we now turn on mass terms, we still want to preserve the $\Gamma_1(4)$
invariance, which means that any new terms that appear should be
$\Gamma_1(4)$ forms with the proper weight.  Assuming that the $m_i$
have weight zero, and insisting on the correct weak coupling behavior
leads to the curve
\eqn\quarticm{
y^2=(f_-x^2+a(\tau)x\sum_im_i+b(\tau)\sum_{i<j}m_im_j-u)^2 +(f_+^2-f_-^2)
\prod_i(x+m_i).
}
Written this way, the curve has singularities near $u=m_i^2$
at weak coupling.   The functions $a(\tau)$ and $b(\tau)$ must
be modular forms of weight two that fall off to zero at weak coupling,
otherwise the singularities will be at the wrong values.  Since
all even forms are generated by $f_+$ and $f_-$, this means that
both $a$ and $b$ are proportional to $f_+-f_-$.  This form is a one instanton
term, hence the terms it multiplies can be at most linear in
each of the masses.  The constant for
$a$ is determined by looking at the Seiberg-Witten differential \APS, 
$\lambda=2x(ydP-Pdx)/(P^2-x^2)$.  This has poles at $x=-m_i$ with residue
equal to $m_i$.  There should also be a pole at infinity whose residue
cancels the other residues.  This requirement leads to
\eqn\aeq{
a(\tau)=-{1\over2}(f_+(\tau)-f_-(\tau)).
}
Determining $b(\tau)$ is harder.  As it so happens, if
\eqn\beq{
b(\tau)=-{1\over4}(f_+(\tau)-f_-(\tau)),
}
then the curve in \quarticm\ is invariant under the parity transformation 
$\tau\to \tau+1/2$, $m_1\to-m_1$.  Unlike the cubic curve in \SWII, this
symmetry is hardly manifest in \quarticm.  
However, if one computes the discriminant
of \quarticm, one finds that it is invariant under this transformation if
$b(\tau)$ satisfies \beq\foot{The actual discriminant takes up over 100
pages of text and takes 4 hours on a 100 mip machine to compute.}

Unfortunately, the higher $SU(n)$ do not have this extra parity symmetry.
However, we note an interesting property of \quarticm\ and the singularities
at strong coupling if $b$ has the form in \beq.  This behavior will generalize.
Suppose we choose $m_1=-m_2=m_3=-m_4=m$, then \quarticm\ reduces to
\eqn\qms{\eqalign{
y^2&=(f_-x^2+{1\over2}(f_+-f_-)m^2-u)^2+(f_+^2-f_-^2)(x^2-m^2)^2\cr
   &=\left[(f_-+f)x^2+({1\over2}(f_+-f_-)-f)m^2-u\right]
     \left[(f_--f)x^2+({1\over2}(f_+-f_-)+f)m^2-u\right],
}}
where $f^2=f_-^2-f_+^2$.  The curve is singular when the roots inside
one set of square brackets match with the roots inside the other set.
This occurs when $u=(f_++f_-)m^2/2$.  At weak coupling $f_\pm\approx1$,
hence the singularity occurs near $u=m^2$.  However, for the strong
coupling limit, $-1/\tau\to i\infty$, $f_++f_-\sim (-i\tau)^2e^{2\pi i\tau}$,
hence the singularity approaches the point $u=0$.  In other words, going
to strong coupling runs the effective mass to zero.  If the coefficient
were different, then at strong coupling we would have found the singularity
to occur at $u\sim (-i\tau)^2m^2$.

\subsec{$\beta$-functions}

Using the curve for the $N_f=4$ case, it is straightforward to compute
the full nonperturbative $\beta$ function for the $N_f=0$ and massless
$N_f=2$ cases.
Although this is outside the main development of the paper, we are unaware
of this calculation appearing previously in the literature, and in
any case will be generalizable to the $SU(3)$ $\beta$-functions.  

The curve for the massless $N_f=4$ case can be expanded to
\eqn\sutwoexp{
y^2=f_+^2x^4-2uf_-x+u^2
}
and the argument $\tau$ of $f_+$ and $f_-$ is the actual coupling.  By 
rescaling
$x$ and $y$ and shifting $\tau$ by 1, one can reexpress the curve as
\eqn\sutwoII{
y^2=x^4-2u'F(\tau)x+u'^2,
}
where $F(\tau)={\th_3^4(2\tau)+\th_1^4(2\tau)\over\th_2^4(2\tau)}$.  We
have replaced $u$ by $u'$ in \sutwoII, in order to distinguish it from
the expectation value $u$ that appears in the scale noninvariant theories.  
$\tau$ does not change when $u'$ is varied.

The curve in the $N_f=0$ case is given by \refs{\AF,\KLTY}
\eqn\sutwonfzero{
y^2=x^4-2ux^2+u^2-\Lambda^4,
}
hence comparing \sutwoII\ with \sutwonfzero, one finds that the coupling
for the $N_f=0$ case satisfies 
\eqn\Fnfzero{
F(\tau)=u(u-\Lambda^4)^{-1/2}.
}
Taking derivatives  with respect to $\Lambda$ on both sides gives
\eqn\gderiv{
\Lambda {d\tau\over d\Lambda}F'(\tau)={2u\Lambda^4\over(u^2-\Lambda^4)^{3/2}}
=2F(F+1)(F-1),}
where $F'$ is the derivative of $F$ with respect to $\tau$.
Hence the $\beta$-function is
\eqn\betazerod{
\beta=\Lambda {d\tau\over d\Lambda}={2F(F+1)(F-1)\over F'}.
}
\betazerod\ can be further reduced by noting that 
$\th_2^4\partial_\tau\th_1^4-\th_1^4\partial_\tau\th_2^4$ is a modular
form of weight six.  Even weight modular forms are generated by $\th_1^4$
and $\th_2^4$, hence this derivative should be a combination of these
functions.  By matching to the leading order behavior and to the transformation
properties under $\tau\to -1/(4\tau)$ we find that
\eqn\derivrel{\th_2^4(2\tau)\partial_\tau\th_1^4(2\tau)-
\th_1^4(2\tau)\partial_\tau\th_2^4(2\tau)=
2\pi i \th_1^4(2\tau)\th_2^4(2\tau)\th_3^4(2\tau).}
Plugging \derivrel\ into \betazerod\ leads to the extremely simple expression
\eqn\betazero{
\beta={2\over\pi i}{\th_3^4(2\tau)+\th_1^4(2\tau)\over\th_2^8(2\tau)}.
}

 From \betazero, it is clear that $\beta$ is a weight negative two modular 
{\it function} 
of $\Gamma_1(4)$, and under the transformation $\tau\to-1/(4\tau)$, $\beta$
transforms as $\beta\to 1/(4\tau^2)\beta$.  We have also verified that the
first few terms in this expansion are consistent with the results in \FP,
where derivatives of  the coordinates $a$ and $a_D$ 
are expressed in terms of elliptic functions.

The $\beta$-function has a zero when $\th_3^4(2\tau)=-\th_1^4(2\tau)$.  
In this case
$\tau=(1+i)/2$, up to a $\Gamma(2)$ transformation.  From \Fnfzero, we
see that this point corresponds to $u=0$, hence it is not surprising
to find a zero of the $\beta$-function since there is now only one scale
in the theory.  The $\beta$-function is also singular as $\th_2(2\tau)$ 
approaches zero which corresponds to the limits $\tau=n$, where $n$
is any integer.  These points are of course where the monopoles and dyons
become massless.

The curve in the massless $N_f=2$ case is given by
\eqn\sutwonftwo{
y^2=x^4-2(u+3\Lambda^2/8)x^2+(u-\Lambda^4/8)^2,
}
hence we have that  
\eqn\Fnftwo{
F(\tau)={u+3\Lambda^2/8\over u-\Lambda^4/8}.
}
Taking derivatives with respect to zero and substituting back in $F$ for
$\Lambda^2/u$ results in
\eqn\betatwo{
\beta=\Lambda {d\tau\over d\Lambda}={(F-1)(F+3)\over 2F'}
={1\over2\pi i}
{\th_3^4(2\tau)+\th_2^4(2\tau)\over\th_3^4(2\tau)\th_2^4(2\tau)}.
}
The $\beta$-function blows up when $\th_2^4$ or $\th_3^4$ approach
zero, corresponding to
massless monopoles or dyons.  There is also a zero when 
$2\th_3^4(2\tau)=\th_1^4(2\tau)$.  This is the coupling if $u=0$.  

In principle, one should be able to compute the $\beta$-function for
$N_f=1,3$ as well.  The standard quartic equation
in these cases has even and odd powers of $x$.  There exists an $SL(2,C)$
transformation into the forms of \sutwonfzero\ and \sutwonftwo, but it is
highly nontrivial.  For massless $N_f=1$, the $\beta$-function can be 
found using
the cubic form of the curves in \SWII.  Compare the curves
\eqn\cubic{
y^2=(x-e_1u')(x-e_2u')(x-e_3u')}
and
\eqn\sutwonfone{
y^2=x^2(x-u)-\Lambda^6/64,
}
where the $e_i$ are given in \SWII.  One finds after shifting $x$ by
a constant in \cubic,
\eqn\betaone{
\beta=\Lambda {d\tau\over d\Lambda}={6(2+F)\over F'},} 
where 
\eqn\Fnfone{
F={(2\th_2^4+\th_1^4)(2\th_1^4+\th_2^4)(\th_2^4-\th_1^4)\over
(\th_1^8+\th_2^8+\th_1^4\th_2^4)^{3/2}}.}

\newsec{$SU(3)$ with $N_f=6$} 

Let $\tau$ be the true coupling when $m_i=0$ and $u=0$.  In \MN, it was
shown that a genus two surface with period matrix $\tau C$ is given
by the hyperelliptic curve
\eqn\genusII{
y^2=(r(\tau)x^3-v)^2+s(\tau)x^6,
}
where 
\eqn\rval{\eqalign{
r(\tau)&={(\vth_1\vth_2\vth_3)^2\over2}(\vth_2^2+\vth_3^2)
(\vth_1^2+\vth_3^2)(\vth_1^2-\vth_2^2)\cr
s(\tau)&={27\over4}\vth_1^8\vth_2^8\vth_3^8.
}}
$\vth_i$ are the genus two theta functions
\eqn\thfn{\eqalign{
\vth_0&=\vth\left[\matrix{0&0\cr0&0}\right](\tau C)\qquad\qquad
\vth_1=\vth\left[\matrix{0&0\cr1&0}\right](\tau C)\cr
\vth_2&=\vth\left[\matrix{1&0\cr0&0}\right](\tau C)\qquad\qquad
\vth_3=\vth\left[\matrix{0&1\cr1&0}\right](\tau C)
}}
If we absorb a factor a factor of $\vth_1\vth_2\vth_3$ into $x$,
then the curve can be rewritten as
\eqn\ngenusII{
y^2=(r'(\tau)x^3-v)^2+s'(\tau)x^6,}
where
\eqn\nrval{\eqalign{
r'(\tau)&={(\vth_2^2+\vth_3^2)
(\vth_1^2+\vth_3^2)(\vth_1^2-\vth_2^2)\over2\vth_1\vth_2\vth_3}\cr
s'(\tau)&={27\over4}\vth_1^2\vth_2^2\vth_3^2.
}}

We expect to be able to rewrite this curve in terms of $\Gamma_1(3)$
forms.  Luckily, these forms can be classified.  Consider the quantities
\eqn\fpmt{\eqalign{
f_\pm(\tau)&=\left({\eta^3(\tau)\over\eta(3\tau)}\right)^3\pm
              \left(3{\eta^3(3\tau)\over\eta(\tau)}\right)^3,\cr
      \eta(\tau)&=q^{1\over24}\prod_{n=1}^\infty(1-q^n).}
}
Both $f_+$ and $f_-$ are modular forms of weight three for $\Gamma_1(3)$.
In fact these forms generate all forms of weight $3m$, where $m$ is
any positive integer.  These forms also transform nicely under
$\tau\to -1/(3\tau)$, with 
$f_\pm\to\pm(3\tau)^3i^{-3}f_\pm$.   
The space of forms of weight one and two are
one dimensional \Sch\ and are generated by $f_1=(f_+)^{1/3}$.  Hence 
 $f_1$ and $f_-$ generate all of the modular forms. 

The functions $r'(\tau)$ and $s'(\tau)$ have very simple relations
to these forms, namely $r'=f_-$ and $s'=f_+^2-f_-^2$.  The curve is
then
\eqn\nngenusII{\eqalign{
y^2&=(f_-x^3-v)^2+(f_+^2-f_-^2)x^6\cr
   &=P(x)^2+(f_+^2-f_-^2)x^6. 
}}
The form of the curve in \nngenusII\ is remarkably similar to the $SU(2)$
curve in \quartic.

We now wish to turn on the other expectation values.  If we keep the
quarks massless and turn on $u$, then at weak coupling $P(x)$ should
approach $P(x)=x^3-ux-v$.  From \nngenusII, we see that $v$ has weight
three under $\Gamma_1(3)$  if $x$ has weight zero, therefore, $ux$ has weight
two.  Thus, $ux$ must be multipied by a weight one form in $P(x)$
so that the curve is $\Gamma_1(3)$ invariant.  The unique form with
the correct weak coupling behavior is $f_1$, hence the generic
massless curve is
\eqn\uvcurve{
y^2=(f_-x^3-f_1ux-v)^2+(f_+^2-f_-^2)x^6. 
}
In terms of the genus two theta functions, $f_1(\tau)=\vth_0(\tau C)$, and
hence this curve matches the curve given previously in \MN\ after
a rescaling in $x$.

At this point the reader might be wondering why there is an $f_-$ in
front of the $x^3$ term instead of $f_+$, since both functions have the same
weak coupling behavior.  It turns out that this is necessary in order to
have the correct duality behavior.  Suppose that $u=0$.  Then, in moving 
from weak
coupling to strong coupling, we expect that quarks will be mapped to 
monopoles and vice versa.  In order for this to happen, the integrals
around the $a$ cycles of the hyperelliptic curve, which correspond to the
electric coordinates $a^I$,
should smoothly go to integrals around the $b$ cycles, which correspond to
the magnetic coordinates
$a^I_D$, when $\tau\to -1/(3\tau)$.  Under this transformation, $f_-$
picks up an extra sign.  
Because of this, if we had chosen the function
in front of the $x^3$ to be $f_+$, then  we would have found that the
$a^I$ map back to themselves under $\tau\to -1/(3\tau)$.  
But with the coefficient $f_-$,  
we find that the $a^I$ transform to the $a^I_D$ under 
$\tau\to -1/(3\tau)$. 

The case with nonzero masses is similar to the situation found for $SU(2)$.
The masses are assumed to have weight zero, hence any mass terms that appear
in $P(x)$ must be multiplied by weight three forms and must fall off to
zero at weak coupling.  Hence these extra terms are proportional to
$f_+-f_-$.  Since this is a one instanton term, they must be at most
linear in each of the individual masses.  Furthermore, there cannot be
a term $u\sum m_i$, since this is a weight two form and hence has
to multiply a weight one-form that falls to zero at weak coupling.  No
such form exists.  Hence the massive curve should be of the form
\eqn\muvcurve{\eqalign{
&y^2=\cr
&\biggl(f_-x^3+(f_+-f_-)
(ax^2\sum m_i+bx\sum_{i<j}m_im_j+c\sum_{i<j<k}m_im_jm_k)
-f_1ux-v\biggr)^2\cr
&\qquad\qquad\qquad+(f_+^2-f_-^2)\prod_i(x+m_i), 
}}
where $a$, $b$ and $c$ are to be determined.  In order to have the correct
residue in $\lambda$ at $x=\infty$, $a$ should be set to $a=-1/2$.
To set $b$ and $c$, we use the argument used in the previous section
for $SU(2)$.  First consider the case $u=0$ and $m_1=e^{2\pi i/3}m_2=
e^{4\pi i/3}m_3=m_4=e^{2\pi i/3}m_5=e^{4\pi i/3}m_6=m$.  Then the curve
in \muvcurve\ reduces to
\eqn\muvred{\eqalign{ 
y^2&=(f_-x^3+2cm^3(f_+-f_-)-v)^2+(f_+^2-f_-^2)(x^3+m^3)^2\cr
&=\left[(f_-+f)x^3+2cm^3(f_+-f_-)+fm^3-v\right]\cr
&\qquad\qquad\times  \left[(f_--f)x^3+2cm^3(f_+-f_-)-fm^3-v\right].
}}
As before, $f^2=f_-^2-f_+^2$.
The roots for a polynomial inside a set of square brackets matches the
roots of the polynomial inside the other set of brackets if
\eqn\rootmatch{
v=m^3(2c(f_+-f_-)-f_-).}
The singularity approaches $v=0$ for strong coupling if $c=-1/4$, otherwise
the singularity would be found at large $v$ for fixed $m$.  To find a suitable
value for $b$, let $v=0$, $m_1=-m_2=m_3=-m_4=m$ and $m_5=m_6=0$.  This curve
is already singular, but an extra singularity arises if 
$u=m^2(f_--2b(f_+-f_-))$.  If $b=-1/4$, then the singularity approaches $u=0$
in the strong coupling limit.  Hence the final curve is
\eqn\fmuvcurve{\eqalign{
y^2=\biggl(f_-x^3-
{f_+-f_-\over4}(2x^2\sum m_i&+x\sum_{i<j}m_im_j+\sum_{i<j<k}m_im_jm_k)
-f_1ux-v\biggr)^2\cr
 &+(f_+^2-f_-^2)\prod_i(x+m_i). 
}}
If we take one of the masses to infinity while taking the coupling to
zero, we can reduce this to a an $N_f=5$ theory which has precisely
the same form as in \HO.

\subsec{$\beta$-functions}

It is also straightforward to find the $\beta$-functions for the $N_f=0,3$
cases, using the  same procedure as in the previous section.  For the $N_f=0$
case with $u=0$, we find
\eqn\betasuthzd{
\beta=\Lambda {d\tau\over d\Lambda}={3F(F+1)(F-1)\over F'},
}
where now, $F=f_-/f_+$.  
Using the fact that 
$f_-\partial_\tau f_+-f_+\partial_\tau f_-$ is a modular form of weight
eight and based on its leading order behavior and transformation properties,
one finds
\eqn\derthrel{
f_-\partial_\tau f_+-f_+\partial_\tau f_-=\pi i(f_+^2-f_-^2)f_1^2.
}
Hence \betasuthzd\ can be reexpressed as
\eqn\betasuthz{
\beta=\Lambda {d\tau\over d\Lambda}={3\over\pi i}{f_-\over f_1^5}  
}

  From \fpmt, one
finds that $\beta$ in \betasuthz\ has a zero
 when $\eta^4(\tau)=3\eta^4(3\tau)$.  Using the fact that
$\eta^2(-1/\tau)=-i\tau\eta^2(\tau)$, it then follows
that $\beta=0$, when $\tau=i/\sqrt{3}$.
The $\beta$-function blows up when $f_1$ approaches zero, which occurs when
$\eta^4(\tau)=-3\eta^4(3\tau)$.  Using the relation $\eta^4(-1/(\tau+1))
=e^{\pi i/3}(i\tau+i)^2\eta^4(\tau)$, we find that $\beta$ diverges if 
$\tau=1/2+i/(2\sqrt{3})$.  This singularity  
occurs at the cusp described in \AF\ and \AD.       

In the massless $N_f=3$ case, by matching curves we find that the $u=0$ 
$\beta$-function is
\eqn\betathree{
\beta=\Lambda {d\tau\over d\Lambda}={3(F-1)(F+3)\over 4F'}=
{3\over4\pi i}{f_-(3f_++f_-)\over f_1^5(f_++f_-)}.
}
This still has a singularity when $\tau=1/2 +i/(2\sqrt{3})$, but it also
is singular if $f_++f_-=0$, which occurs at $\tau=n$.

An interesting check of $\beta$ in \betasuthz\ and \betathree
 would be to compute the higher
instanton corrections to the coupling.  Work on this is in progress.    
At this time, we do not know the $\beta$-functions for other values of $N_f$.

\newsec{$SU(n)$, $N_f=2n$}

For the $SU(n)$ groups with $n>3$, two problems arise.  The first problem
is that the dimension of $\Gamma_1(n)$ forms of low weight is somewhat
large.  This makes it difficult to choose coefficients based on uniqueness
arguments.  

The second problem is that there is no region in the space
of expectation values where the true quantum coupling is proportional to
the matrix $C$ in \cc.  This basically means that the parameter $\tau$
that appears in the curve will not be the actual coupling for any choice
of expectation values.

We have yet to overcome these problems, but we describe some of the issues
involved.
To understand the relation of $\tau$ to the coupling, let us consider
the case where all bare masses are zero and all expectation values are
zero except for the casimir $s_n$.  One can calculate the perturbative
quantum corrections
to the coupling, giving
\eqn\qucoup{
T_{qu}=\tau C+{i \over\pi}G,
}
where the entries of the matrix $G$ are given by
\eqn\quentry{G_{mm}=\log\left(4n^2\sin^2{m\pi\over n}\right)\qquad\qquad
G_{ml}=\log\left(-2ni{\sin{m\pi\over n}\sin {l\pi\over n}
\over\sin{|l-m|\pi\over n}}\right).
}
The $\log 2n$ and $\log i$ terms can be absorbed into the classical coupling,
however the log sine terms cannot be absorbed, otherwise the coupling
won't have the proper behavior under Weyl reflections.  From \quentry\ 
it is clear
that for $n>3$, the full coupling is not proportional to the Cartan matrix
$C$.  

Let us concentrate on the $SU(4)$ and $SU(5)$ cases.  For $SU(4)$, we can
rewrite the coupling as
\eqn\sufour{
T=\tau C+\eps B,\qquad\qquad B=\left(\matrix{0&-2&2\cr-2&-4&-2\cr2&-2&0}\right)
,}
where in weak coupling,
$\eps={i\over2\pi}\log\sin{\pi\over4}-{i\over2\pi}\log\sin{2\pi\over4}
=-{i\over4\pi}\log2$.  Under the transformation $T\to T(8C^{-1}T+1)^{-1}$, $T$
transforms to
\eqn\Ttran{
T\to \widetilde\tau C+\widetilde\eps B
}
where
\eqn\tildeeq{
\widetilde\tau={\tau+8\tau^2-32\eps^2\over(1+8(\tau+2\eps))(1+8(\tau-2\eps))}
\qquad\qquad\widetilde\eps={\eps\over(1+8(\tau+2\eps))(1+8(\tau-2\eps))}.
}
Letting $\tau_1=\tau+2\eps$ and  $\tau_2=\tau-2\eps$ leads to 
the transformations
\eqn\newtautr{
\tau_1\to{\tau_1\over 1+8\tau_1}\qquad\qquad\tau_2\to{\tau_2\over 1+8\tau_2}.
}

A similar situation exists for $SU(5)$.  Here we can write the coupling
matrix as
\eqn\sufive{
T=\tau C+\eps B,\qquad\qquad B=\left(\matrix{2&-1&1&3\cr-1&-2&-3&1\cr
1&-3&-2&-1\cr3&1&-1&2}\right)
,}
where in this case, 
$\eps={I\over2\pi}(\log\sin{\pi\over5}-\log\sin{2\pi\over5})$ for weak 
coupling.  We can then define $\tau_1=\tau+\sqrt{5}\eps$ and 
$\tau_2=\tau-\sqrt{5}\eps$, which transform as  
\eqn\newtautrfi{
\tau_1\to{\tau_1\over 1+5\tau_1}\qquad\qquad\tau_2\to{\tau_2\over 1+5\tau_2},
}
under $T\to T(5C^{-1}T+1)^{-1}$.
For higher groups, the same sort of procedure can be followed, but instead
of one or two parameters that transform under $\Gamma_1(n)$ or $\Gamma_1(2n)$,
there are $(n-1)/2$ ($n/2$) parameters for $n$ odd (even).

The natural generalization of the massless $SU(2)$ and $SU(3)$ cases is to
assume that the hyperelliptic curve is of the form
\eqn\hypersun{
y^2=(f_-(\tau)x^n-\sum_{i=2}^n s_n (f_+(\tau))^{1-i/n}x^{n-i})^2+
(f_+^2(\tau)-f_-^2(\tau))x^2n,
}
where $f_+$ and $f_-$ are $\Gamma_1(n)$ ($\Gamma_1(2n)$ forms of weight $n$
for $n$ odd (even).
But here is where the  two difficulties arise that need to be overcome.  
First, it is not
clear how $\tau$ should be chosen.  For instance for $SU(4)$ (or $SU(5)$), 
there
are two variables, $\tau_1$ and $\tau_2$, that transform under $\Gamma_1(8)$ 
(or $\Gamma_1(5)$).  Second, the dimensions of the forms $f_-$ and $f_+$ are
greater than 1.  Hence, uniqueness arguments are not sufficient for
determing the true equation.  

{\it Note added:}  As this paper was being typed, a preprint appeared \AY\ 
that has some overlap with the discussion in section 2. 

{\bf Acknowledgements}: We thank Nick Warner for many helpful discussions.  
This research
was supported in part by D.O.E.~grant DE-FG03-84ER-40168.


\listrefs

\bye